\documentclass[a4paper,fleqn,]{cas-sc}

\usepackage{graphicx}   
\usepackage{tabularx}   
\usepackage{booktabs}  
\usepackage{url}        
\usepackage{xspace}     
\usepackage{etoolbox}   

\usepackage{caption}
\usepackage{float}
\usepackage[numbers]{natbib}
\usepackage{multicol}
\usepackage{capt-of}
\usepackage{subcaption}

\DeclareCaptionFont{cas}{\sffamily\small}

\def\tsc#1{\csdef{#1}{\textsc{\lowercase{#1}}\xspace}}
\tsc{WGM}
\tsc{QE}

\begin{document}
\let\WriteBookmarks\relax
\def\floatpagepagefraction{1}
\def\textpagefraction{.001}

\shorttitle{}
\title [mode = title]{Virtual Reality User Interface Design: Best Practices and Implementation}

\author[1]{Esin Mehmedova}[orcid=0009-0001-1948-2316]
\credit{Methodology, Writing, Surveys, Data Collection}

\affiliation[1]{organization={Technical University of Munich},
            city={Heilbronn},
            country={Germany}}

\author[1]{Santiago Berrezueta-Guzman}[orcid=0000-0001-5559-2056]
\ead{s.berrezueta@tum.de}
\credit{Conceptualization, Writing, Supervision}

\author[1]{Stefan Wagner}[orcid=0000-0002-5256-8429]
\credit{Conceptualization, Writing, Supervision}

\begin{abstract}
Designing effective user interfaces (UIs) for virtual reality (VR) is essential to enhance user immersion, usability, comfort, and accessibility in virtual environments. Despite the growing adoption of VR across domains, there is a noticeable lack of unified and comprehensive design guidelines for VR UI design. To address this gap, we conducted a systematic literature review to identify existing best practices and propose 28 unified guidelines for UI development in VR.

Building on these insights, this research proposes a framework to guide the creation of more effective VR interfaces. To demonstrate and validate these practices, we developed a VR application called \textit{FlUId} and an interactive \textit{Web Tool} that serves as a guideline explorer and project planning resource for developers. A user study was conducted to evaluate the impact of the proposed guidelines. The findings aim to bridge the gap between theory and practice, offering concrete recommendations and digital tools for VR designers and developers.
\end{abstract}



\maketitle
\begin{multicols}{2}

\section{Introduction}\label{I}

Virtual Reality (VR) is a computer-generated simulation of a three-dimensional environment that users can interact with and explore in an immersive, often headset-driven, experience. VR originated in the 1960s \cite{hamdanindustry}. It utilizes head-mounted displays (HMDs) to create the illusion of an alternate reality, immersing users in interactive digital environments \cite{byam2019best}. A well-designed VR experience can enhance user engagement, satisfaction, and immersion \cite{saranya2024development}, but excessive use of the technology may also lead to various health problems and harm personal development \cite{puspitasari2022review}. This highlights the importance of thoughtful design, as interface elements and usability directly influence users' mental models and decisions, with user needs and experiences serving as critical design criteria \cite{chen2024survey}.

While gesture-based interface components continue to emerge, traditional user interface (UI) elements remain popular to this day \cite{yeo2024entering}. However, for a truly immersive VR experience, it is essential to explore different UI design approaches that support a more natural user interaction \cite{kharoub20193d}. Current methods for designing virtual 3-dimensional (3D) interfaces are typically based on direct adaptations of the desktop metaphor and the related "window, icon, menu, pointing device" (WIMP) paradigm, resulting in a user experience (UX) gap between physical and virtual reality \cite{bermejo2021exploring}. 

This adaptation is problematic because interaction in 3D virtual environments involves mapping traditional 2-dimensional (2D) input devices to 3D space, which limits how content is perceived when 3D elements are shown as 2D rendered images \cite{caputo2019gestural}. Unlike conventional 2D interfaces used on desktops and mobile devices, VR environments do not have fixed UI placement boundaries, making traditional UI design principles less directly applicable \cite{pandey2024pilot}. 

The current lack of established design guidelines for VR interfaces leads to significant limitations, particularly in selecting appropriate interaction modes for immersive experiences \cite{chen2024survey}. Existing VR interaction models struggle with user fatigue, motion sickness, and the overall difficulty of designing interfaces that ensure usability and user well-being \cite{zhou2023design, khanvirtual}. There is a lack of established principles for displaying text, with open questions about optimal contrast ratios and other visual parameters affecting readability \cite{kojic2020user}. Additionally, recent studies highlight the challenges in improving menu placement and interaction techniques in VR, emphasizing the need for careful evaluation to enhance usability, accuracy, and user comfort \cite{byam2019best, wang2021experimental}.

To address all these challenges and more, this research conducts a comprehensive literature review to identify current best practices and guidelines for VR UI design. Based on this analysis, we compiled a set of 28 best practices and policies that synthesize existing research findings into unified design recommendations for VR interfaces, accessible to any VR designer or developer on a Wiki page on GitHub\footnote{VR-UI-Guidelines: \url{https://github.com/TUM-HN/VR-UI-Guidelines/wiki}}. To validate and demonstrate these guidelines, we developed a VR application called \textit{Fl\textbf{UI}d}\footnote{VR-UI-Guidelines: \url{https://github.com/TUM-HN/VR-UI-Guidelines}} that showcases both good and bad UI design principles, providing users with a direct comparison of different design approaches. We also conducted a user study to evaluate the effectiveness of these guidelines and their impact on user experience in VR environments. 

This work contributes by providing a unified framework of design guidelines that can help designers create more effective, comfortable, and accessible VR UIs. The findings aim to bridge the gap between theoretical research and practical application, offering concrete guidance for improving user experience in virtual reality environments.

\subsection{Applications of Virtual Reality}
The application of VR technology spans a wide range of domains, each presenting unique UI design considerations. 

Immersive learning environments are used in education to create engaging and interactive experiences \cite{sobchyshak2025}. The study \cite{gonzalez202310} proposed a set of best practices for immersive learning design, emphasizing realistic interaction within virtual environments. The work by Xie \cite{Xie} complemented this by identifying key design points that shape user interface design in educational VR applications, particularly focusing on information clarity, interaction flow, and spatial orientation. The research by Zhou et al. \cite{zhou2023design} presented a set of 3D user interface design paradigms tailored for VR exhibitions to enhance usability and immersion in virtual environments.

Beyond education, VR is widely adopted in gaming and entertainment. Wang et al. \cite{wang2021authenticity} explored design strategies that support collaboration, interactivity, and authenticity in VR games, highlighting how these principles enhance user experience. Works like Larocco \cite{larocco2020developing}, Alves et al. \cite{alves2020evaluation}, and Çatak et al. \cite{ccatak2020guideline} analyzed existing industry best practices and manufacturer guidelines, like Oculus and Unity, to propose frameworks that can be generalized across different VR platforms. Kelly et al. \cite{kelly2022design} presented a VR application designed to support mindfulness practices, emphasizing user comfort, customizable environments, guidance levels, and adaptable interaction modes to foster mindful awareness in safe, controlled settings.

VR interfaces facilitate treatment, recovery, and user well-being in healthcare and rehabilitation \cite{berrezuetaguzman2025}. Ramaseri Chandra et al. \cite{ramaseri2022systematic} conducted a systematic survey addressing the causes of cybersickness, a recurring issue in VR environments, and compiled guidelines for improving the virtual experience by minimizing discomfort. Veličković and Milovanović \cite{velivckovic2021improvement} examined specific design factors in VR rehabilitation, demonstrating that user navigation approaches are the primary cause of cybersickness. Yahaya et al. \cite{yahaya2022comparative} further contributed by addressing cybersickness reduction strategies in immersive VR applications, particularly within road safety education for children.

Accessibility and inclusivity are also key themes in current VR research. Zhang et al. \cite{zhang2025inclusive} derived a set of design guidelines to support inclusive avatar design for people with disabilities, covering avatar appearance, body dynamics, assistive technologies, peripherals, and customization control. Ha et al. \cite{ha2022novel} developed a calibration-free hybrid brain-computer interface system that enables hands-free virtual reality control, particularly targeting users with limited mobility. Ijaz et al. \cite{ijaz2022design} identified a set of design recommendations for interfaces that accommodate older adults. Kalaiselvi et al. \cite{kalaiselvi} investigated gesture elicitation in VR for low-vision users, presenting an alternative interaction method that improves immersion, satisfaction, and completion times.

\subsection{User Interface Design in Virtual Reality}
Current methods for designing virtual 3D interfaces are typically based on direct adaptations of the desktop metaphor and the related "window, icon, menu, pointing device" (WIMP) paradigm, resulting in a user experience gap between physical and virtual reality \cite{bermejo2021exploring}. This adaptation is problematic because interaction in 3D virtual environments involves mapping traditional 2D input devices to 3D space, which limits how content is perceived when 3D elements are shown as 2D rendered images \cite{caputo2019gestural}. VR introduces distinct challenges and factors such as field of view, user orientation, absence of a defined reference plane, and depth-related complexities that differ significantly from flat 2D displays, which need to be considered \cite{kaur2019exploring}.

Previous research has examined different types of user interfaces, such as graphical user interfaces (GUIs) and natural user interfaces (NUIs), particularly in immersive environments. Regazzoni et al. \cite{regazzoni2018virtual} presented guidelines for designing NUIs in VR, focusing on usability, ergonomics, and gesture interaction. The work of Cardona-Reyes et al. \cite{cardona2020natural} built on these guidelines to support therapy in children with ADHD. Similarly to NUIs, three-dimensional GUIs lack comprehensive design heuristics, and understanding the cognitive workload of such systems requires considering different variables \cite{kaur2020cognitive}. Kaur and Yammiyavar \cite{kaur2017comparative} compared 2D and 3D interface elements to propose guidelines for 3D GUIs.

Research has focused on the design of interface elements like menus, buttons, and keyboards, as well as text readability in VR. Various menu designs, such as Open Palm Menu \cite{azai2018open}, Tap-tap Menu \cite{azai2018tap}, and an accordion-style
menu \cite{kwon2017spatial}, demonstrate alternative placement options for menus in VR as well as different interaction types. Moreover, the innovative 3D hexa-ring menu introduced by Pandey and Sorathia \cite{pandey2024pilot} showcases a three-dimensional approach that enhances user immersion and ease of interaction, outperforming the traditional 2D design. Kim and Xiong \cite{kim2022pseudo} introduced pseudo-haptic features for button design that simulate feedback in VR environments, improving user experience and satisfaction. Dudley et al. \cite{dudley2019performance} compared various keyboard typing strategies in VR, concluding that using physical surfaces and two-finger interaction leads to faster typing performance. Dingler et al. \cite{dingler2018vr} investigated factors such as angular height, vergence distance, view box parameters, and text font and background, and identified user preferences for better text readability in VR.

\subsection{Interaction and Navigation in Virtual Reality}
A crucial aspect of UI design in VR is the selection of effective interaction mechanisms and navigation strategies, which significantly influence user experience.
Laine and Suk \cite{laine2024investigating} investigated how gesture-based interaction affects presence in VR by identifying potential presence factors and proposing a framework for designing and evaluating gesture-based immersive environments.
The work by Byam \cite{byam2019best} compared three common VR interaction techniques, such as virtual laser pointer, tapping, and d-pad selection, evaluating user performance and preference, and concluded that virtual laser pointer and tapping are the recommended methods for VR selection tasks. Moreover, Kim et al. \cite{kim2017study} proposed and evaluated a gaze pointer-based user interface for mobile virtual reality, demonstrating how different interface designs, based on visual field range, feedback systems, information transfer methods, and background color, can influence user immersion, presence, convenience, and the risk of VR sickness.

Movement and navigation techniques are also crucial for the overall user experience in VR. Cenydd and Headleand \cite{ap2018movement} provided an overview of diverse locomotion methods, including teleportation, gamepad locomotion, and other position-tracking methods, and presented design recommendations for movement in virtual reality, particularly within gaming environments. Otaran and Farkhatdinov \cite{otaran2024exploring} explored user preferences for various VR walking interfaces, revealing three preferred VR navigation methods: joystick, point-and-teleport, and walking in place. Byam’s research \cite{byam2019best} further explored player navigation strategies, highlighting teleportation as the preferred method over joystick navigation among users \cite{byam2019best}. 
Moreover, Bektas et al. \cite{bektacs2021systematic} introduced the Limbic Chair, an embodied VR interface that enables independent leg movements, compared to a gamepad. The results showed potential for reducing simulator sickness and enhancing immersion in specific VR scenarios despite slower performance and higher workload.

All the reviewed literature above highlights the importance of VR UI design in enhancing immersion, usability, comfort, and accessibility across a wide range of VR applications. Although various application types, interface components, and interaction techniques have been explored, there remains a lack of unified, comprehensive guidelines that synthesize these findings into well-defined best practices. This paper provides a unified framework of design guidelines that can help designers create more effective, comfortable, and accessible VR interfaces, based on the exhaustive analysis of literature and a proof of concept by a VR app that verifies the reliability of these guidelines. 

\section{Methodology}\label{M}
We adopted a multi-step methodological approach that included a systematic literature review to identify and synthesize existing guidelines, the formulation of a categorized set of 28 design recommendations, and their practical implementation in a VR prototype application. To evaluate the usability and effectiveness of these guidelines, we conducted a user study comparing interfaces designed with and without adherence to the proposed best practices.

\subsection{Literature Review}
This process involved systematically collecting and analyzing research papers focusing on VR user interface design elements and interaction techniques. The primary search platform was Google Scholar. Through this process, an initial set of 131 papers published between 2015 and 2025 was identified.

To ensure the relevance of the selected literature, studies were excluded if they were not in English, primarily focused on augmented reality (AR), mixed reality (MR), or extended reality (XR), or did not address the user interface aspects of virtual reality. This screening process resulted in a selection of 97 papers.

Each of these 97 papers was then examined in detail. During this phase, an additional 42 papers were excluded due to their lack of relevant experimental insight. This process resulted in a final set of 55 papers that formed the analytical basis of this study. The chosen papers were carefully reviewed to identify common themes, design considerations, and established guidelines related to VR UI components and user interaction methods. 
The full selection process is illustrated in Figure~\ref{fig:review}.

\begin{center}
\includegraphics[width=1\columnwidth]{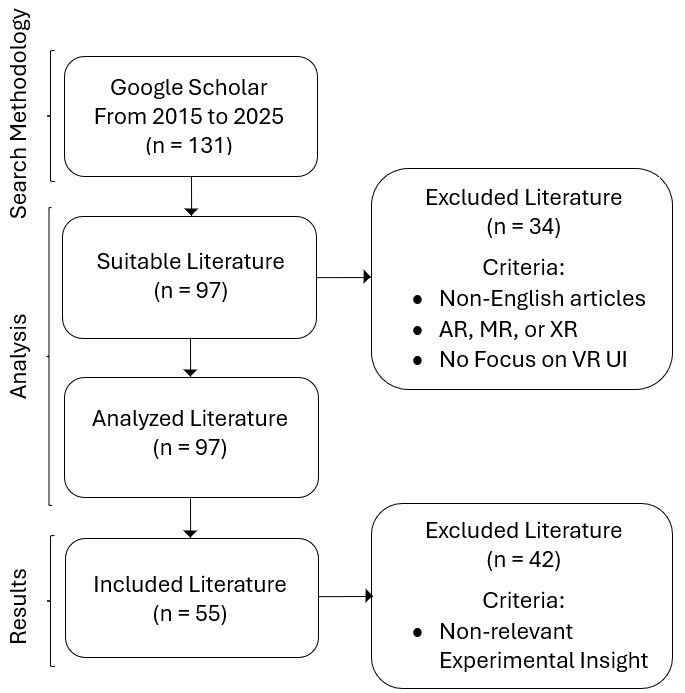}
\captionof{figure}{Literature review process}
\label{fig:review}
\end{center}

\subsection{Best Practices and Guidelines Development}

Building upon the insights gathered from the systematic literature review, a set of 28 detailed guidelines was synthesized in a wiki page on GitHub to provide recommendations for designing effective, usable, and immersive VR interfaces. 
The guidelines are organized into five categories reflecting key design areas in VR environments:

\subsubsection{General VR Design \& UI/UX Principles}
This category encompasses 10 guidelines, featuring manufacturer guidelines from leading VR platforms like Oculus, Leap Motion, and Unity, along with fundamental principles for natural user interfaces, ergonomic design, and 3D graphical user interfaces. 

Alves et al. \cite{alves2020evaluation} compare manufacturer guidelines, emphasizing depth cues, comfortable placement of elements, and integration of interface elements into virtual environments. LaRocco \cite{larocco2020developing} presents the Oculus best practices guide, addressing general user experience considerations including session length management, visual display principles, movement input, and more. Regazzoni et al. \cite{regazzoni2018virtual} present natural user interface design guidelines focusing on dialogue principles, ergonomic considerations, and gesture-based interactions for VR environments. 

Hari Chandana et al. \cite{chandana2023exploring} suggest user experience design best practices in VR, emphasizing the importance of understanding user needs, rapid prototyping, precise wayfinding mechanisms, and intuitive interaction design. Lima et al. \cite{lima2024construction} focus on ergonomics and user experience in VR interface design, highlighting the need to move beyond 2D design principles, design elements in an intuitive way, and incorporate storytelling techniques. Khan et al. \cite{khanvirtual} provide human factors and ergonomic design guidelines in VR, covering performance requirements, accessibility considerations, and visual fatigue mitigation strategies. Kaur and Yammiyavar \cite{kaur2017comparative} offer insights into adaptive interface design, incorporating feedback and cognitive load management. 

Shoikova and Peshev \cite{shoikova2017best} establish best practices for VR design, emphasizing the importance of creating believable environments, stimulating users' senses, and designing avatar-centric interfaces that account for personal space and multitasking. Wall \cite{wall2021empirical} presents empirical findings on VR menu interaction and design, providing recommendations for visualization, haptic feedback, ray-casting, and pre-attentive processing. Yeo et al. \cite{yeo2024entering} offer design recommendations for aligning input device dimensionality with virtual tasks, limiting unnecessary motion, and highlighting the importance of context-dependent interaction.

\subsubsection{VR Game Design \& Development}
This category contains 7 guidelines specifically tailored for VR game environments, covering game-specific considerations including universal design principles for accessibility, Unity-specific best practices, movement mechanics that maintain user comfort, avatar design for inclusive representation, and strategies for minimizing addiction in virtual environments. 

Çatak et al. \cite{ccatak2020guideline} provide comprehensive guidelines for VR game design, covering perceptual design principles such as visual attraction points, as well as depth and sensory cue utilization. Their work also addresses interactional and navigational design guidelines. Dombrowski et al. \cite{dombrowski2019designing} focus on inclusive VR game design through the pillars of universal design, aiming to ensure equitable use for users with diverse abilities and preferences, as well as immersive and accessible content that minimizes physical effort. 

Anuyahong \cite{Anuyahong_Pengnate_2023} demonstrates that there is a strong relationship between effective UI design and positive user experiences while emphasizing the importance of intuitive navigation and consistent visual design. Wang et al. \cite{wang2021authenticity} examine authenticity, interactivity, and collaboration in VR games, providing best practices for creating engaging experiences that support learning and collaborative problem-solving in games. Ap Cenydd and Headleand \cite{ap2018movement} present movement modalities in VR, offering guidelines for player guidance, circular environment design, and comfort preservation through appropriate character distancing and exploration rewards. 

Zhang et al. \cite{zhang2025inclusive} contribute specialized guidelines for inclusive avatar design that support disability representation in social VR environments, addressing avatar body appearance, dynamics, assistive technology integration, and interface considerations for creating accessible and inclusive virtual experiences. Puspitasari and Lee \cite{puspitasari2022review} review persuasive user interface strategies for addressing addiction in virtual environments, providing guidelines for implementing break encouragement systems, user statistics tracking, and adaptive aesthetics to promote responsible VR use.

\subsubsection{Onboarding \& Learning}
This category includes 2 guidelines focused on effectively introducing new users to VR interfaces and designing immersive educational experiences that leverage VR's unique capabilities for embodied learning.

Chauvergne et al. \cite{chauvergne2023user} investigate practices in VR user onboarding, emphasizing the importance of instruction availability, corrective feedback, and context-based learning approaches that integrate instructions into meaningful tasks. Their research highlights the need to allocate time for familiarizing users with the controls before interacting with the virtual environment. Gonzalez-Argote and Gonzalez-Argote \cite{gonzalez202310} establish best practices for immersive learning design, focusing on aligning immersive learning experiences with clear learning objectives, selecting appropriate technology tailored to the target audience, and designing interactive, engaging environments that support meaningful exploration, collaboration, realistic simulations, and progress tracking through feedback and assessment.

\subsubsection{Menus \& Interface Elements}
This category encompasses 7 guidelines addressing the practical aspects of VR interface design, including optimal menu placement strategies that use natural hand movements, hand-adaptive UI systems, selection techniques, and text readability principles for virtual environments.

Azai et al. \cite{azai2018open} introduce the Open Palm Menu concept, providing guidelines for positioning menu items close to users' hands and designing vertical menu layouts for optimal selection efficiency. Li et al. \cite{li2024investigating} offer recommendations for customizable menu configurations and body-area-specific icon placement based on application type. Lou et al. \cite{lou2021hand} present hand-adaptive user interface best practices that emphasize placing interactive objects at lower heights to reduce fatigue and leveraging spatial memory for intuitive object positioning. Lediaeva and Jr \cite{lediaeva2020evaluation} provide specific recommendations for arm-mounted, waist-level, hand-based, and spatial menus with corresponding selection techniques. 

Wentzel et al. \cite{wentzel2024comparison} compare virtual reality menu archetypes, offering design recommendations across different menu layouts, item counts, and levels of hierarchy. Wang et al. \cite{wang2021experimental} investigate menu selection methods in immersive virtual environments, comparing fixed and handheld menu approaches, and provide guidelines for choosing appropriate selection techniques. Dingler et al. \cite{dingler2018vr} assess text parameters for reading in VR, providing specific guidelines for angular height, vergence distance, viewbox parameters, and font selection to optimize text readability and user comfort in virtual environments.

\subsubsection{Cybersickness \& User Comfort}
This category contains 2 guidelines that focus on critical health and safety considerations, providing research-based recommendations for minimizing simulation sickness, reducing visual fatigue, and maintaining user comfort during extended VR sessions.

Ramaseri Chandra et al. \cite{ramaseri2022systematic} provide design guidelines that address latency requirements, movement realism, visual flicker prevention, logical environment design, optimal field of view, and session duration. Yahaya et al. \cite{yahaya2022comparative} present cybersickness reduction guidelines, offering specific recommendations for duration management, user experience preparation, physical setting optimization, movement control strategies and more.

\subsection{Integration}
To support the practical adoption of VR UI/UX best practices, we define four levels of integration, each representing the degree to which a given guideline should be integrated into a VR application. This helps developers prioritize implementation based on usability and impact. Each level is associated with selected guidelines from the developed set, chosen based on their significance, influence on usability and comfort, and ease of implementation.

\textbf{1. Full Integration:} These are critical guidelines that must always be present in a VR application to ensure basic usability, comfort, and safety. They directly influence user interaction.

\textit{Examples:}
\begin{itemize}
    \item \textbf{Guideline 1.4:} Avoid placing content in peripheral areas. Restrict content that requires more time to focus on the center of vision, while content that requires less attention can be arranged in the peripheral areas of vision.
    \item \textbf{Guideline 1.9:} Ensure interactive elements are adequately spaced and sized to make the interaction easy and avoid accidental selections.
     \item \textbf{Guideline 2 (Vision):} Maintain comfortable viewing distances.
    \item \textbf{Guideline 3.2 (Graphic Interface):} The user interface should be simple, functional, and not distracting.
    \item \textbf{Guideline 4.3:} Design for intuitive and ergonomic interactions that enhance user experience.
    \item \textbf{Guideline 5.3:} Position interactive elements within easy, natural reach, particularly in the lower field of view, to enhance user comfort.
\end{itemize}

\textbf{2. Medium Integration:} These guidelines strongly enhance the user experience and accessibility, but may be adapted slightly based on context or platform.

\textit{Examples:}
\begin{itemize}
    \item \textbf{Guideline 1.11:} Avoid pinning UIs to the user’s view to prevent discomfort; if necessary, implement delayed following behavior.
    \item \textbf{Guideline 2 (User Input):} Accommodate left/right-handed users.
    \item \textbf{Guideline 7.2:} Incorporate multiple feedback methods.
    \item \textbf{Guideline 27.1.6:} Design for short exposure duration. Allow the user to pause, rest, and continue later.
\end{itemize}

\textbf{3. Low Integration:} Guidelines in this category add polish, accessibility, or style consistency. They are beneficial, but not essential for the core VR experience.

\textit{Examples:}
\begin{itemize}
    \item \textbf{Guideline 3.3:} Gestures based on real-world actions are generally more intuitive and thus easier to learn.
    \item \textbf{Guideline 6.3.6:} High frequency of color changes and color temperature affect fatigue. Avoid colors with frequent and sudden changes, do not use highly saturated colors. Choose colors with low luminance to minimize fatigue.
    \item \textbf{Guideline 12.4:} Perceptible Information - Use lighting, spatial cues, audio, haptic feedback, and subtitles to ensure content is immersive and accessible to all users.
    \item \textbf{Guideline 27.1.2:} Movement should be realistic. Fast falling, rolling, waveform motion, flipping, and rapid zoom should be avoided. Jumping instead of continuous walking can reduce sickness.
\end{itemize}

\textbf{4. No Integration:} These are optional or context-specific enhancements. They are useful in specialized applications but can be omitted without significant negative impact on basic usability.

\textit{Examples:}
\begin{itemize}
    \item \textbf{Guideline 1.3:} Integrate GUI elements into the virtual environment or character. For example, weapon selection by grabbing a virtual backpack or holster.
    \item \textbf{Guideline 4.6:} Design objects to reflect real-world physics, lighting, and spatial alignment to increase realism and user immersion.
    \item \textbf{Guideline 11.1.6:} Provide consistency for sensory cues. For example, gunshot animation with matching sound or haptic feedback
    \item \textbf{Guideline 16.4.1:} Provide suitable icons, logos, and slogans that reflect disability communities. Users should be able to add well-known icons and slogans to their avatars that represent different disability groups and affirm their identity.
    \item \textbf{Guideline 17.9:} Implement a time limit for application usage.
\end{itemize}

\subsection{VR UI Guidelines Web Tool}\label{WA}
To support practical adoption of the synthesized VR UI design guidelines, we developed a web application\footnote{Web application: \url{https://vr-guidelines.se.cit.tum.de/}} that serves as an interactive guideline explorer and planning tool. While the complete guideline set is available on the project’s GitHub Wiki, the web tool provides a structured workflow for discovering, selecting, and documenting guidelines that are most relevant to a specific VR project.

The tool is implemented using HTML, CSS, and JavaScript. The guideline set can be explored through an interactive filtering interface, as shown in Figure~\ref{fig:web_filter}. Users specify the application context and may additionally select application features, design aspects, design goals and priorities, and design challenges and concerns. Based on these selections, the tool returns a tailored subset of guidelines by matching the chosen categories against each guideline’s assigned tags. This allows developers and designers to quickly narrow the guidelines and focus on recommendations aligned with their intended use case (e.g., education or gaming) and design priorities (e.g., immersion or accessibility). For transparency and exploration, the tool also provides access to the complete list of guidelines without filtering.

\begin{center}
\includegraphics[width=1\columnwidth]{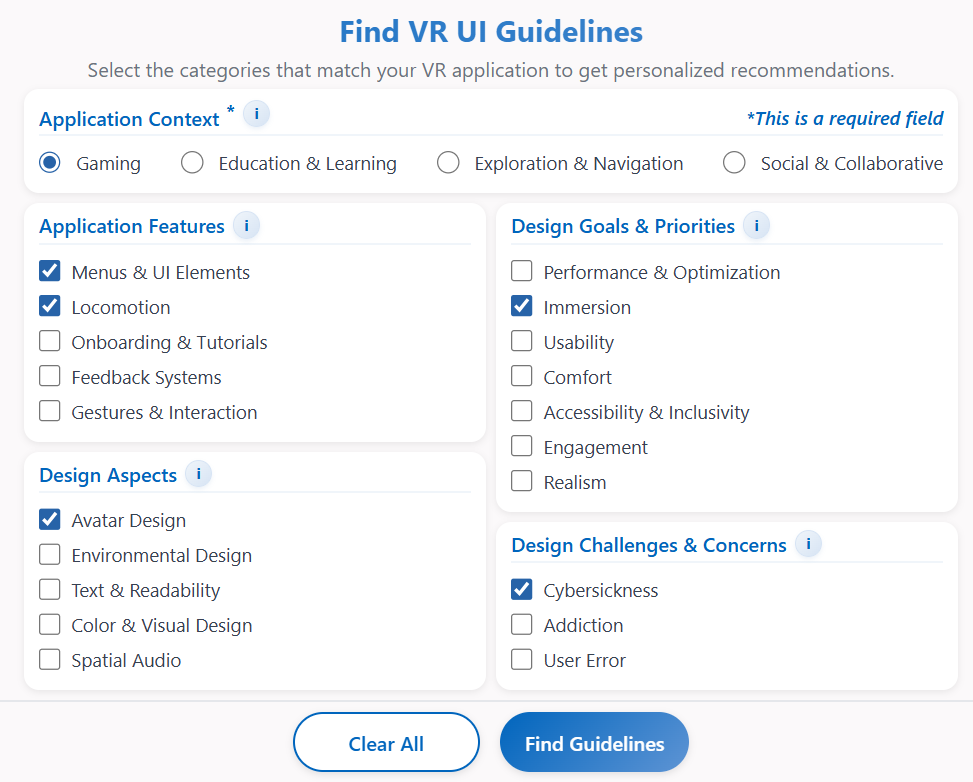}
\captionof{figure}{Filtering interface of the VR UI Guidelines tool. Users select application context and optional categories to obtain a tailored guideline subset.}
\label{fig:web_filter}
\end{center}

To facilitate systematic implementation in the developement process, the web tool includes a checklist where users can mark guidelines as completed as they are integrated into an application (Figure~\ref{fig:web_checklist}). 

\begin{center}
\includegraphics[width=1\columnwidth]{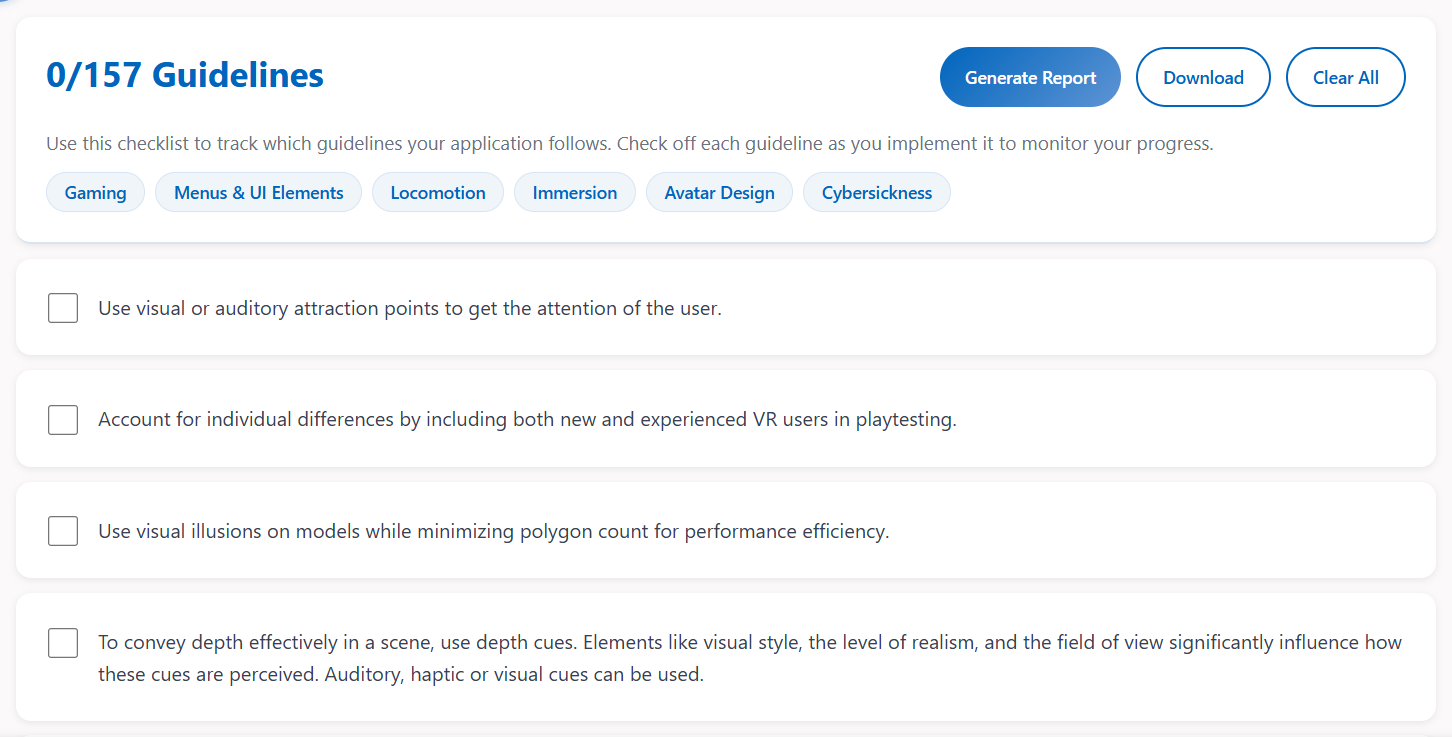}
\captionof{figure}{Checklist for tracking guideline implementation.}
\label{fig:web_checklist}
\end{center}

Building on this, the tool generates an implementation report summarizing overall completion percentage, counts of completed and remaining guidelines, and gives a category-level progress breakdown. This is shown in Figure~\ref{fig:web_report}. Both the checklist and the report can be exported as PDF documents, enabling teams to use the tool as documentation for design decisions, implementation progress, and internal reviews.

Overall, the website extends the contribution of this work from a static set of recommendations to an actionable, interactive tool that supports learning, project-specific guideline selection, and traceable implementation of best practices in VR UI design.

\begin{center}
\includegraphics[width=1\columnwidth]{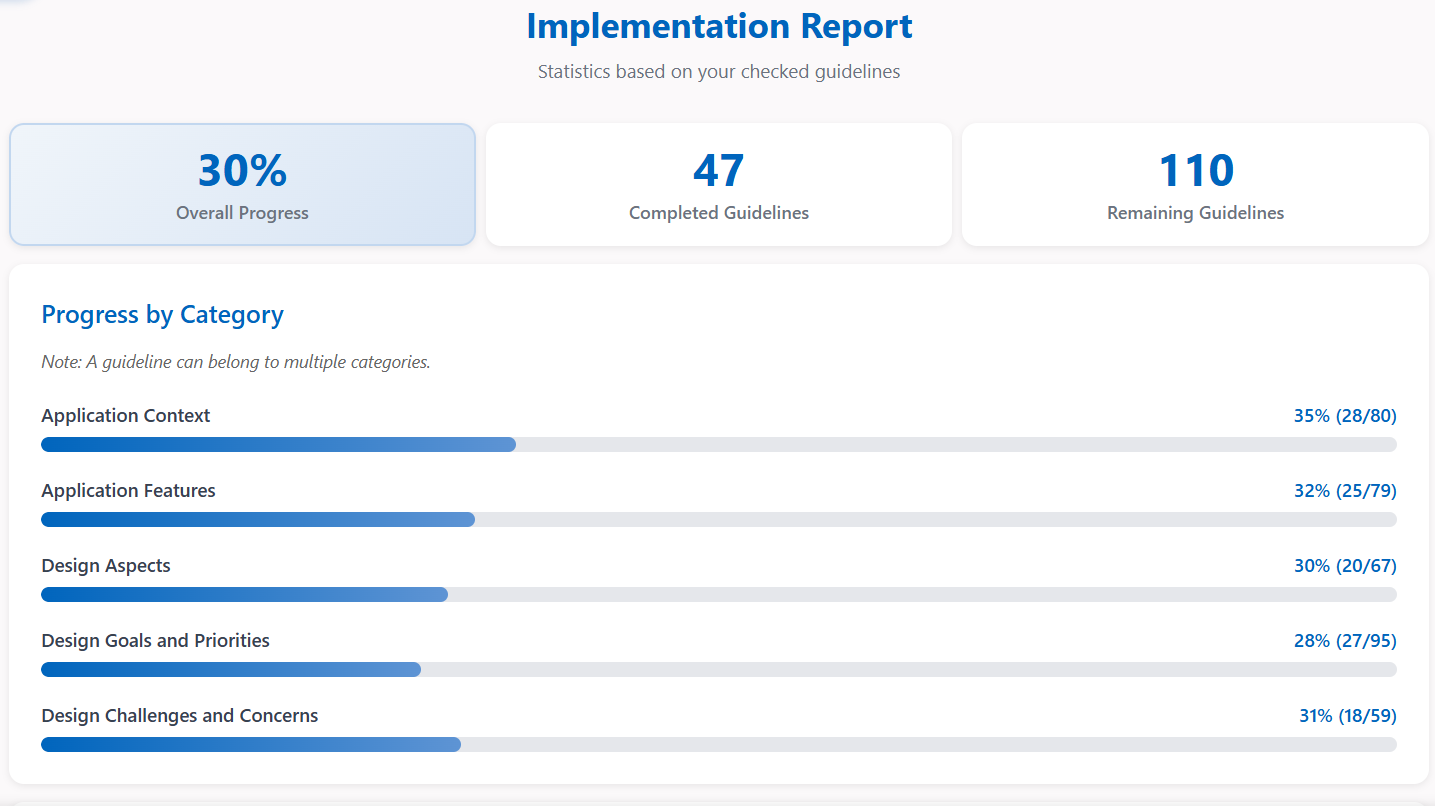}
\captionof{figure}{Automatically generated progress report summarizing overall and category-level completion.}
\label{fig:web_report}
\end{center}

\subsection{Implementation}
To illustrate the effectiveness of the proposed VR UI design guidelines, a virtual reality application titled \textit{FlUId} was developed using the Unity game engine (version 2022.3.62f1). Designed for the Meta Quest 3S headset and Touch Plus controllers, the application serves as an interactive tool for showcasing the impact of both well-designed and poorly designed VR interfaces. It allows users to compare good and bad UI implementations side by side in an immersive environment, highlighting how design decisions influence usability, comfort, and overall user experience.

The application consists of three core components:

\subsubsection{Design One} 
A user settings menu built according to the best practices. Users can interact with the interface by controller ray-cast. Selection happens by pressing the front trigger button of the controller and the joystick is used for scrolling. As shown in Figure~\ref{fig:gooduimenu}, key features of the interface include a comfortable viewing distance, content centered within the user's field of view, and interactive elements such as buttons and toggles placed at lower heights for ergonomic access. The interface provides visual, audio, and haptic feedback, uses vertical placement for interface elements, and includes instructions that remain accessible at all times. It maintains a consistent color scheme, features easily readable text, and avoids blur, floating elements, or irregular shapes.
\begin{figure*}[t]
  \centering
  \begin{minipage}[b]{0.24\textwidth}
    \includegraphics[width=\linewidth]{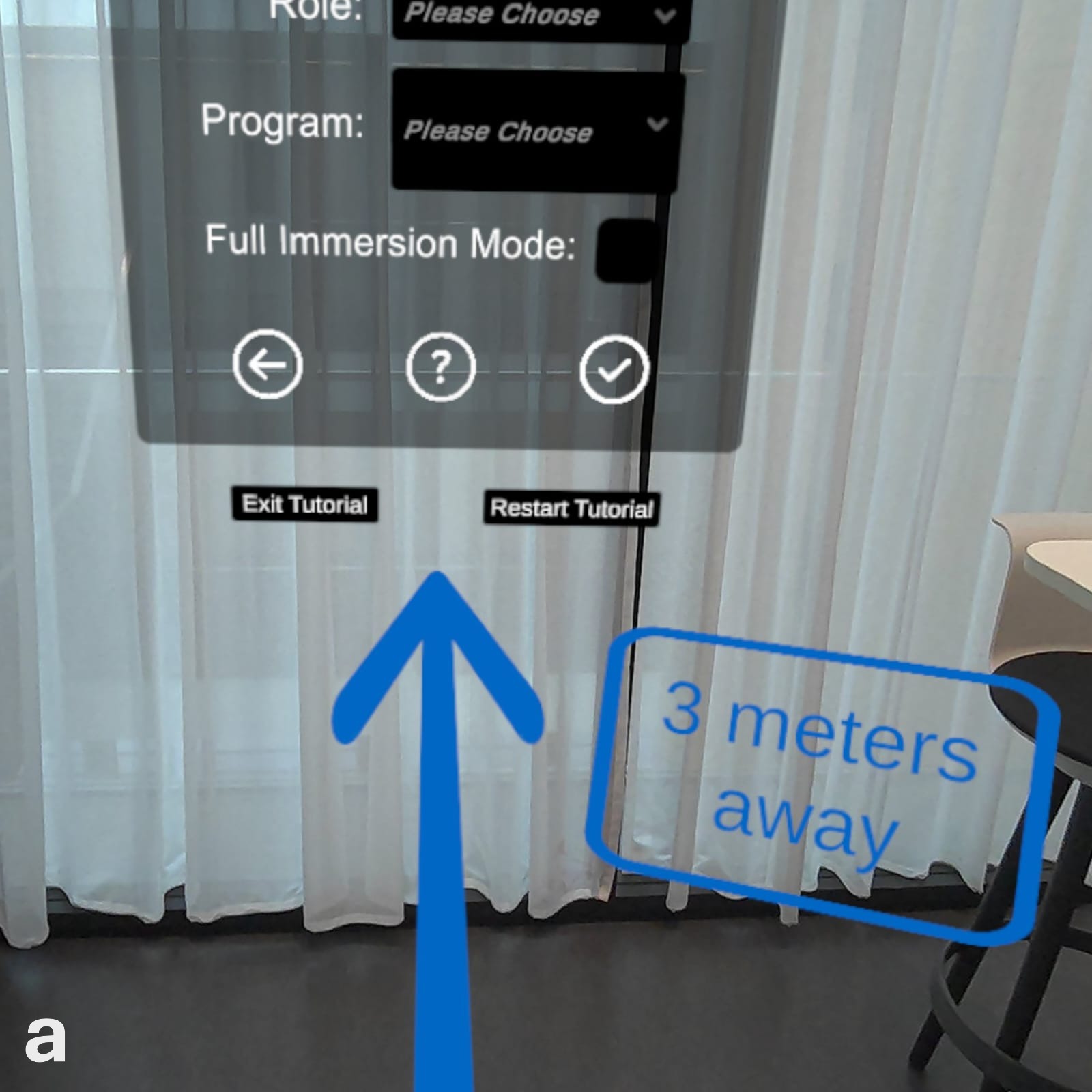}
  \end{minipage}
  \hfill
  \begin{minipage}[b]{0.24\textwidth}
    \includegraphics[width=\linewidth]{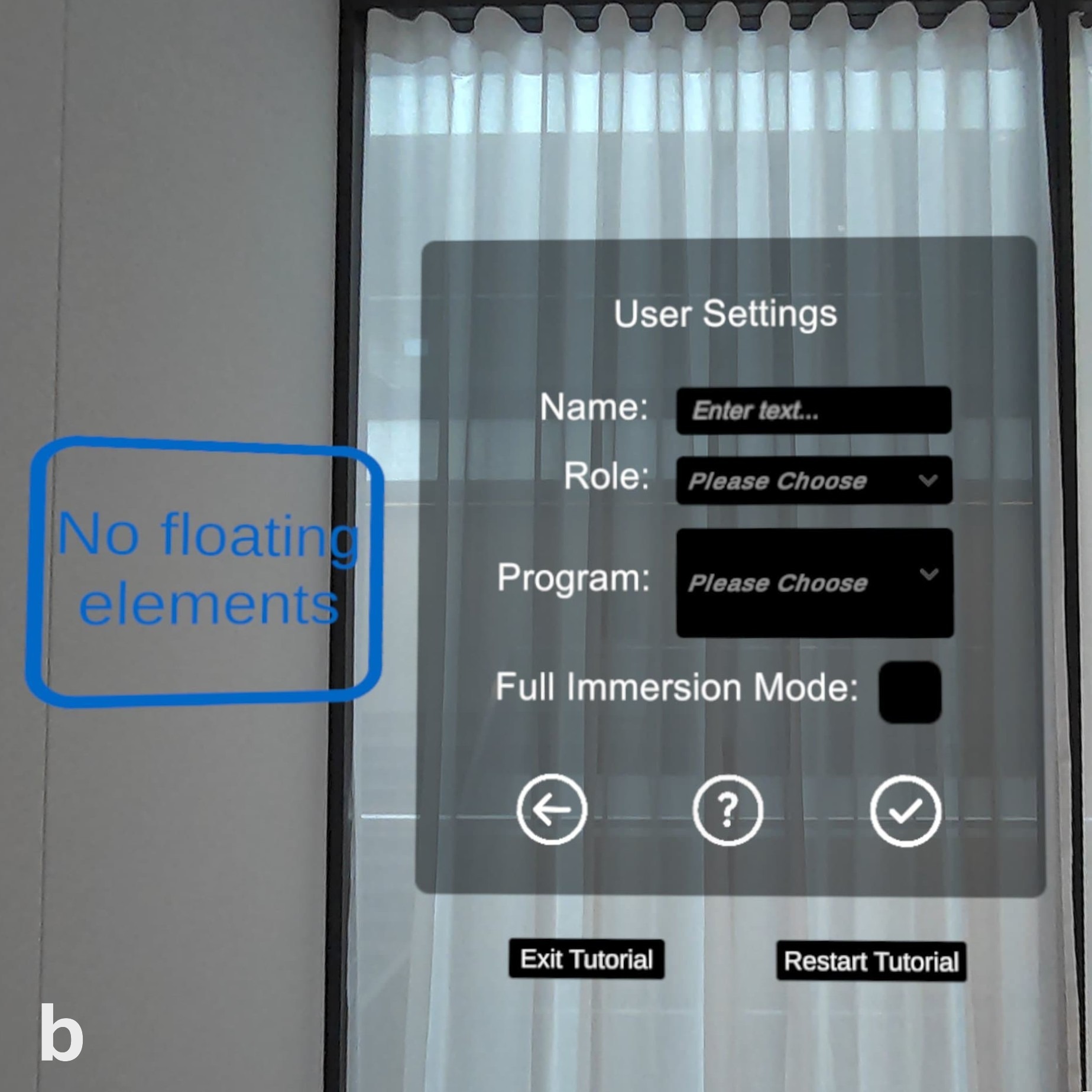}
  \end{minipage}
  \hfill
  \begin{minipage}[b]{0.24\textwidth}
    \includegraphics[width=\linewidth]{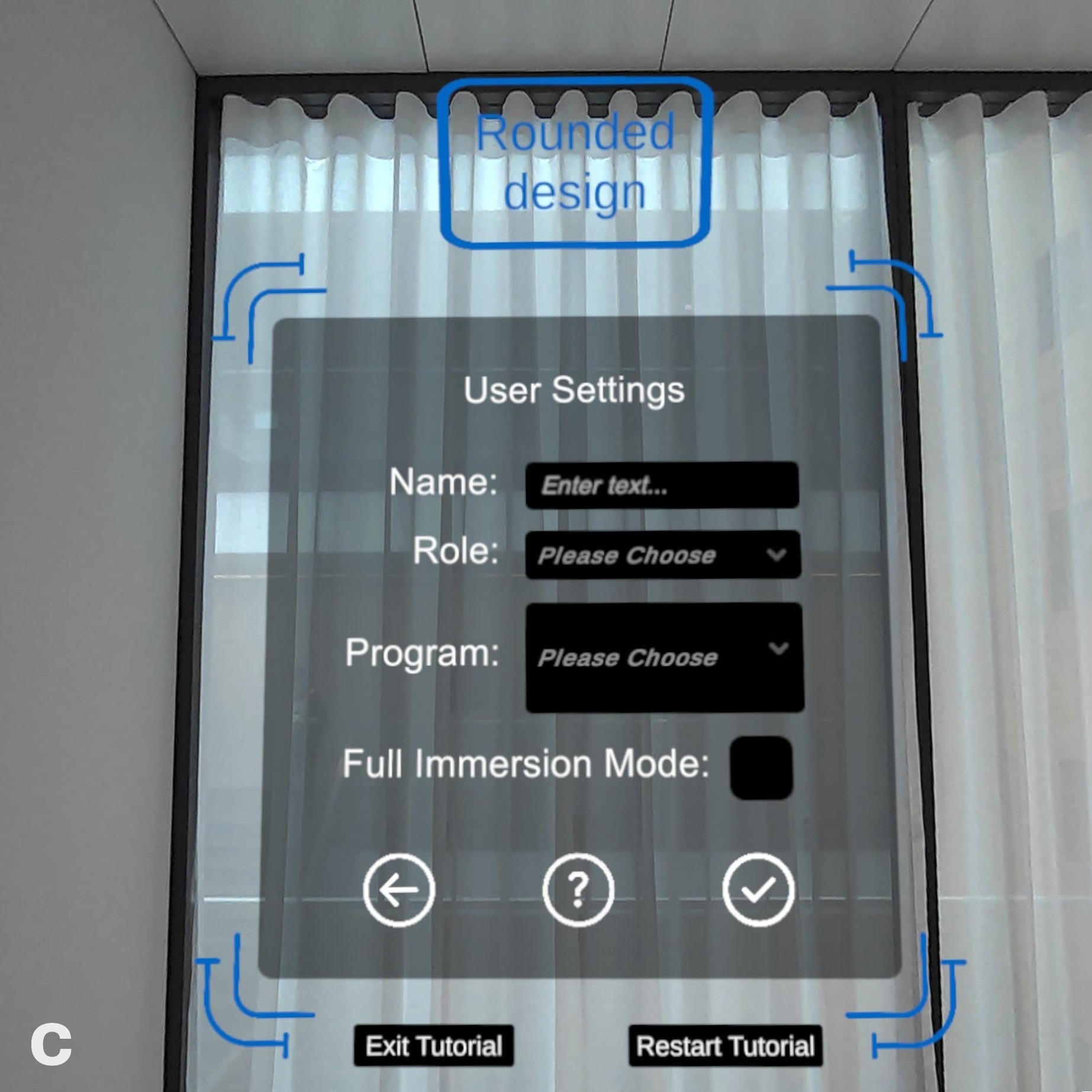}
  \end{minipage}
  \hfill
  \begin{minipage}[b]{0.24\textwidth}
    \includegraphics[width=\linewidth]{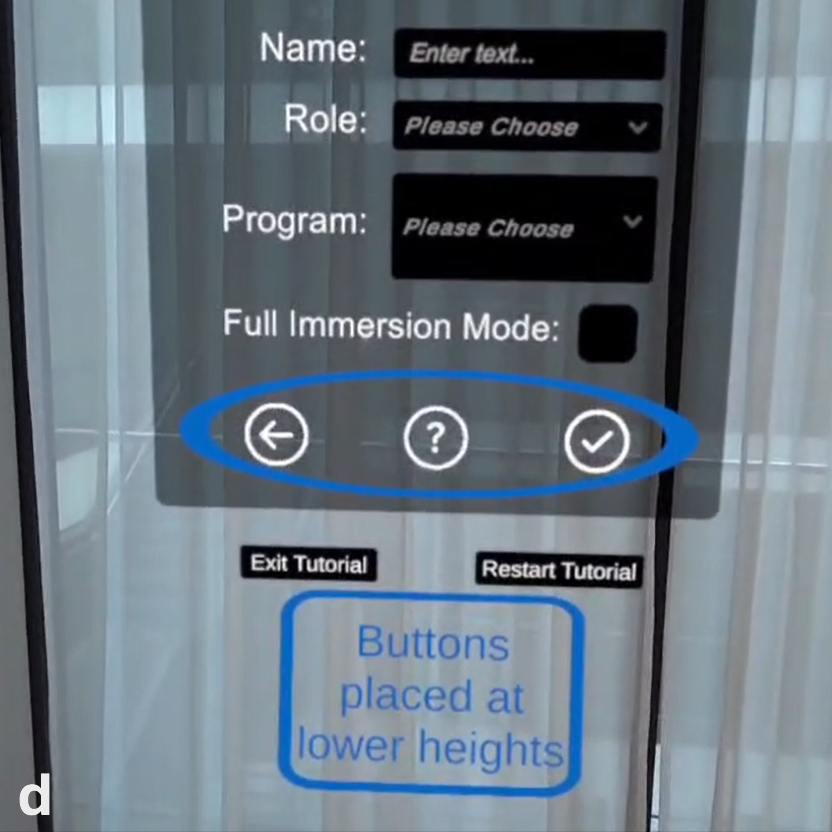}
  \end{minipage}

  \caption{Tutorial highlights showing different UI design choices: (a) distance placement, (b) no floating elements, (c) rounded visuals, and (d) lower button placement.}
  \label{fig:tutorial}
\end{figure*}

\begin{center}
\includegraphics[width=1\columnwidth]{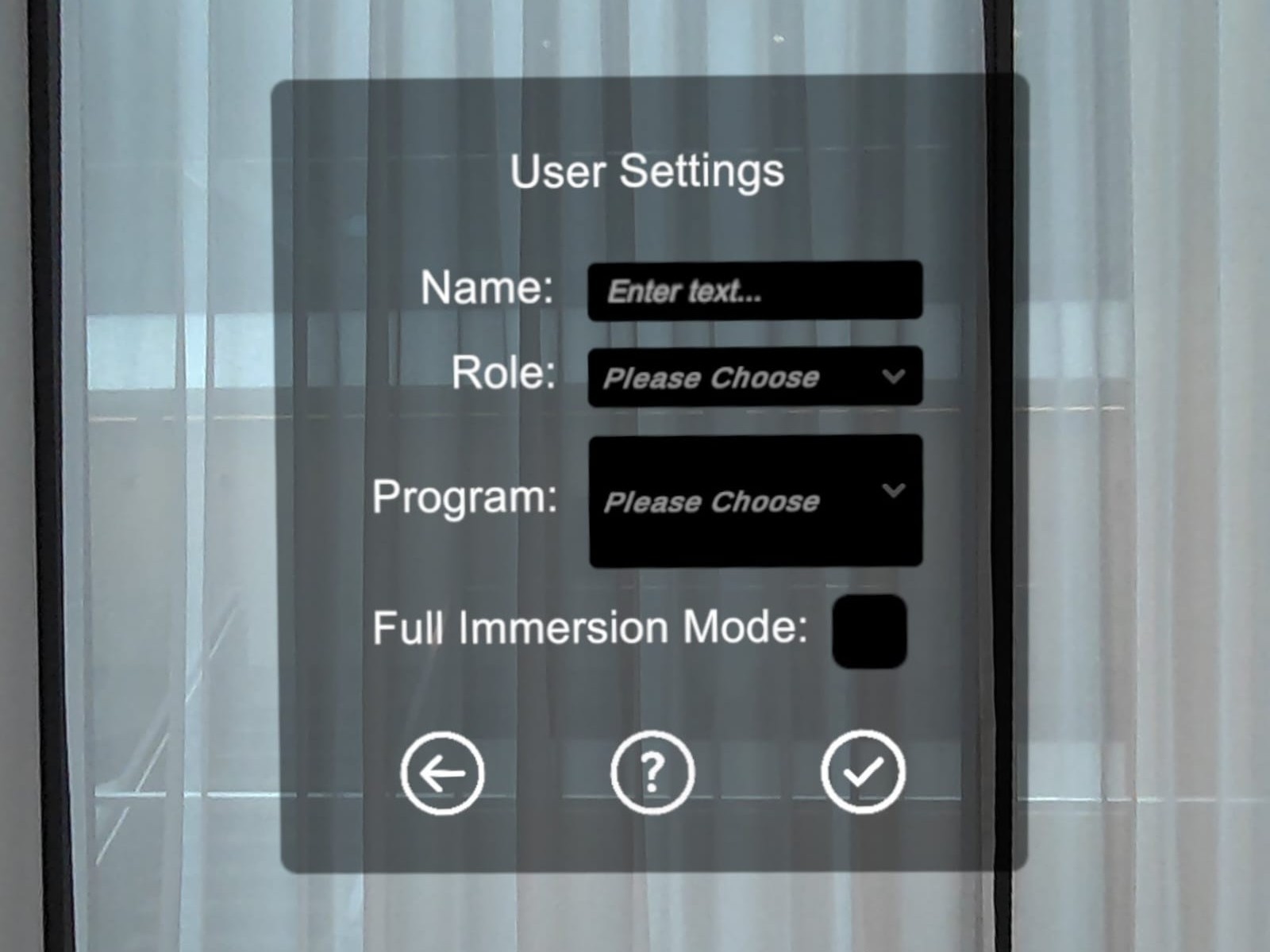}
\captionof{figure}{Design one: A settings menu built according to our VR UI guidelines.}
\label{fig:gooduimenu}
\end{center}

\subsubsection{Design Two}
A user settings menu designed to intentionally violate several of the proposed VR UI design guidelines. Interaction is handled through hand-tracking ray-cast, with selection performed by pinching the thumb and index finger, and scrolling achieved through a pinch-and-drag gesture. As demonstrated in Figure~\ref{fig:baduimenu}, in this interface, elements are misaligned, placed too close to the user, and positioned in their peripheral vision. Feedback is minimal or inconsistent, with no haptic feedback provided. 

No instructions are provided to guide the user, and the menu is arranged horizontally. Text elements violate readability principles, such as appropriate font size, spacing, and background contrast. Additionally, the interface includes floating or blurry elements, as well as components with unbalanced or changing background colors and irregular shapes. The purpose is to create a noticeable contrast in user experience between the two designs.

\begin{center}
\includegraphics[width=1\columnwidth]{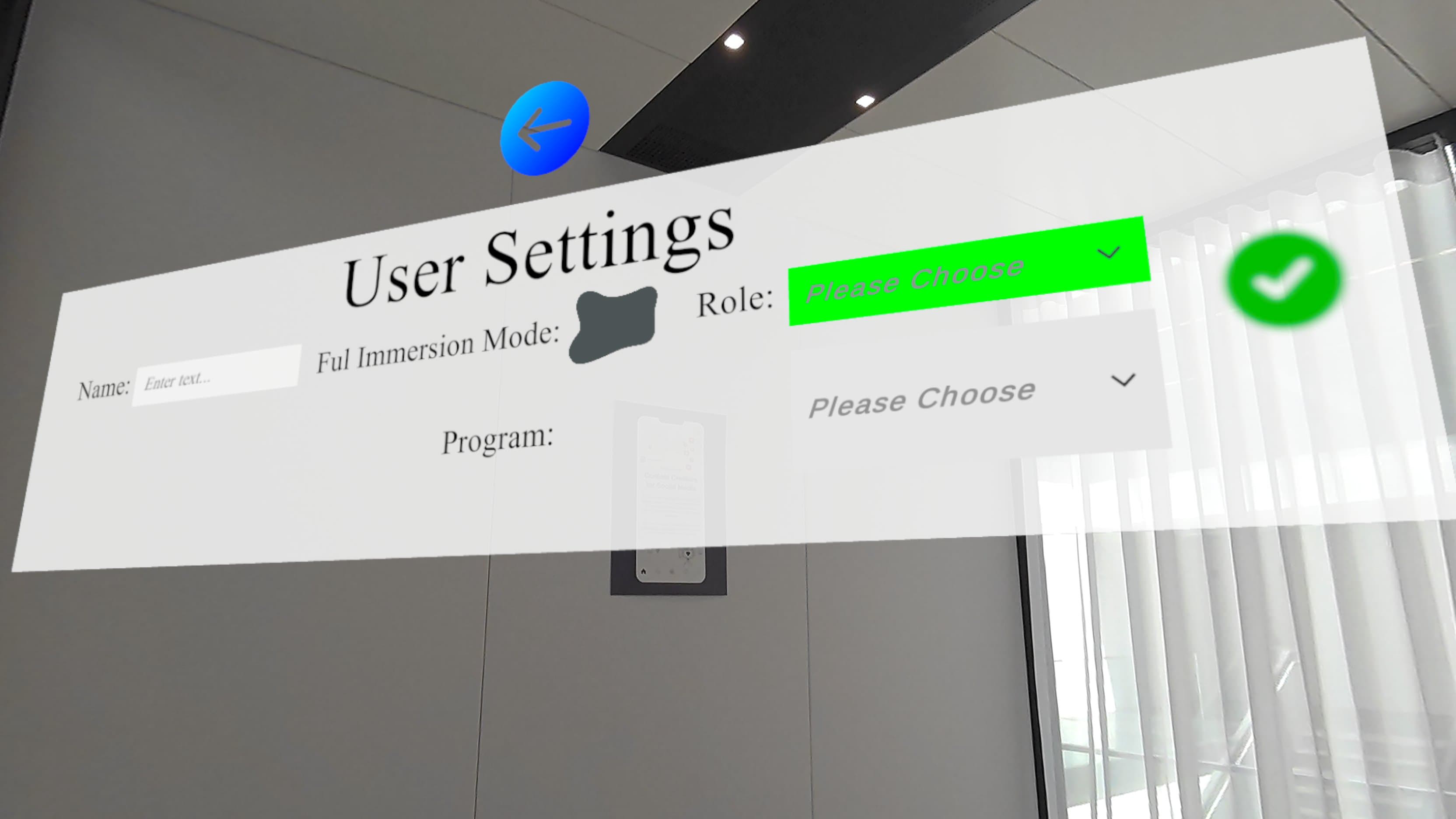}
\captionof{figure}{Design Two: A deliberately poorly designed settings menu that does not adhere to the guidelines.}
\label{fig:baduimenu}
\end{center}

\begin{table*}[h]
  \caption{Survey questions used to evaluate each interface}
  \label{tab:surveyquestions}
  \begin{tabularx}{\textwidth}{@{}cX@{}}
    \toprule
    No. & Survey Question \\
    \midrule
    1  & How intuitive did the user interface feel? \\
    2  & How comfortable was it to use the interface? \\
    3  & How easy was it to interact with the interface elements? \\
    4  & How well organized were the interface elements? \\
    5  & How easy was it to recognize interactive elements (e.g., buttons, toggles)? \\
    6  & How satisfied were you with the feedback provided during interaction? \\
    7  & How visually appealing was the user interface? \\
    8  & How comfortable was the viewing distance of the interface? \\
    9  & How comfortable were the color choices? \\
    10 & How distracting did the visual effects feel? (1 = Very distracting, 5 = Not distracting at all) \\
    11 & How easy and comfortable was it to read the text in the interface? \\
    12 & How would you rate your overall experience? \\
    \bottomrule
  \end{tabularx}
\end{table*}

\subsubsection{Tutorial}
An automatic tutorial is provided to showcase the good design principles applied in \textit{Design One}. It offers a walkthrough of the implemented guidelines, highlighting decisions related to viewing distance, font selection, feedback mechanisms, colors used, and element placement. Figure~\ref{fig:tutorial} shows parts of the tutorial, which demonstrate the distance, visual design, and position of the elements.

\subsection{User Study}
To evaluate the effectiveness of the application and the implementation of VR UI design guidelines, a user study was conducted with 20 participants. 

The participants were divided into two equal groups:
\begin{itemize}
    \item Group A (10 participants) began with \textit{Design One}, interacted with the interface, and then filled out a survey about their experience. Afterward, they interacted with \textit{Design Two} and completed the same survey again.
    \item Group B (10 participants) followed the reverse order, starting with \textit{Design Two}, followed by \textit{Design One}, with a survey completed after each.
\end{itemize}

\begin{center}
\includegraphics[width=1\columnwidth]{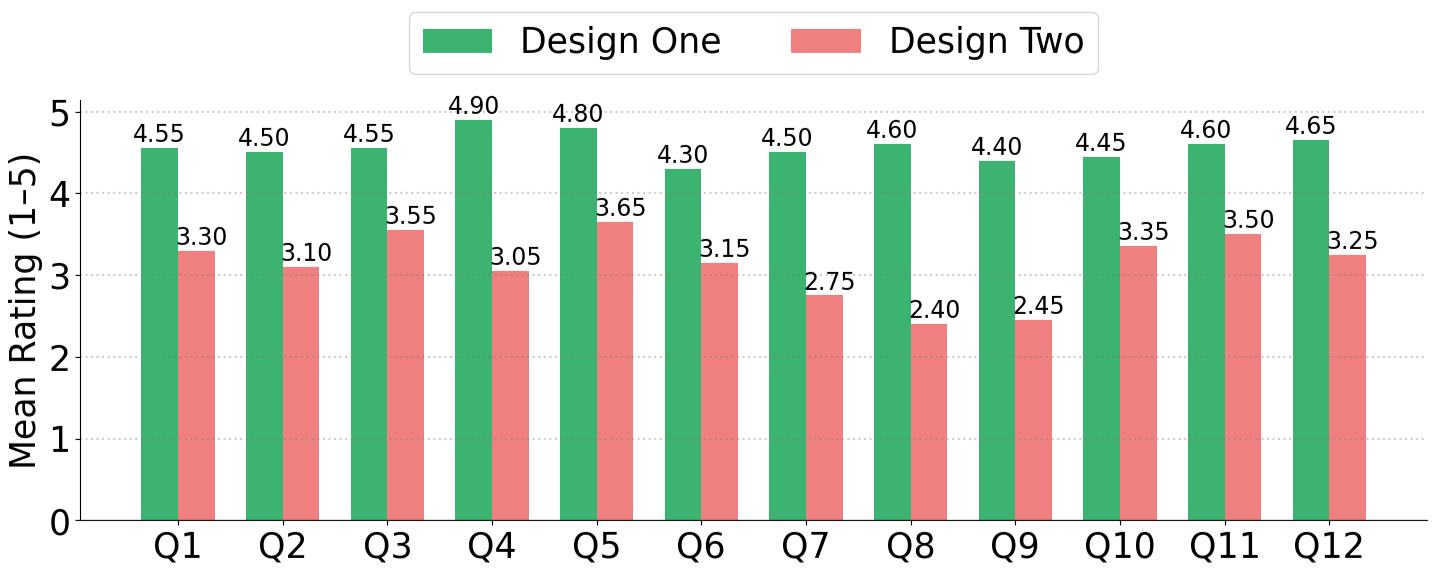}
\captionof{figure}{Mean Likert scores (1–5) for each of the 12 evaluation questions, comparing Design One and Design Two.}
\label{fig:survey}
\end{center}
Before beginning the experiment, all participants were given a brief demonstration on how to put on the headset and use the controllers and hand positioning to interact with the interfaces. This ensured that all participants were familiar with the interaction mechanics before evaluating the designs. Participants were not informed which of the two designs was intended to represent good or bad design. This was done to ensure unbiased feedback based solely on their experience. At the beginning of each interaction, participants were instructed to freely explore the design and form an opinion about what they liked and disliked.

The survey consisted of 12 Likert-scale questions rated from 1 to 5 (worst to best) assessing aspects such as usability, comfort, clarity, ease of interaction, and overall experience. In addition, participants were provided with an open-ended field where they could freely write any observations, thoughts, or preferences regarding each design. After completing both interactions and surveys, participants were asked to choose which design they preferred overall.
The full list of survey questions is provided in Table~\ref{tab:surveyquestions}.

\section{Results}\label{R}
The survey responses were analyzed to assess the usability, comfort, clarity, and overall experience between the two designs, while also examining how the sequence in which participants encountered each design affected their evaluations.

\subsection{Quantitative Analysis}
\textit{Design One} consistently received higher ratings across all questions in both exposure orders. Figure~\ref{fig:survey} presents the mean ratings for each survey question, comparing the performance of \textit{Design One} and \textit{Design Two}. The values represent the combined average scores for each design, based on responses from both Group A and Group B, regardless of whether the design was experienced first or second.

There is a notable drop in scores for \textit{Design Two} compared to \textit{Design One} across all twelve survey questions. The most significant differences appear in Q2, Q4, Q7, Q8, and Q9, which correspond to comfort, organization of interface elements, visual appeal, viewing distance, and color choices, respectively. These areas highlight key weaknesses in the poorly designed interface, particularly in terms of ergonomics and visual appeal. Additionally, overall user experience (Q12) differed significantly between the two designs, with an average rating of 4.65 for \textit{Design One} and 3.25 for \textit{Design Two}. While interaction ease (Q3) and feedback satisfaction (Q6) also showed noticeable gaps, the difference was less pronounced, suggesting that users were more tolerant of differences in interaction methods (controller vs. hand-tracking) than poor visual design. This suggests that users preferred clarity and comfort over innovative or complex interaction modalities.
\end{multicols}

\begin{multicols}{2}
When experienced first with Group A, \textit{Design One} averaged a rating of 4.6 out of 5 across all questions. With Group B, it maintained a similar high performance with a slightly lower average rating of 4.5, indicating a consistently positive user experience, even after initial exposure to a less optimal interface. Figure~\ref{fig:designone} shows a slight decline in Q6 (Feedback), Q7 (Visual Appeal), and Q9 (Color Comfort). This can be attributed to some users either not noticing the haptic feedback or finding it unremarkable, as well as a few comments suggesting the design felt visually plain or lacking color. Q11 (Text Readability) shows a noticeable increase when \textit{Design One} was experienced second, indicating that users in Group B found it easier to read, especially in contrast to the less readable text in \textit{Design Two}.

\begin{center}
\includegraphics[width=1\columnwidth]{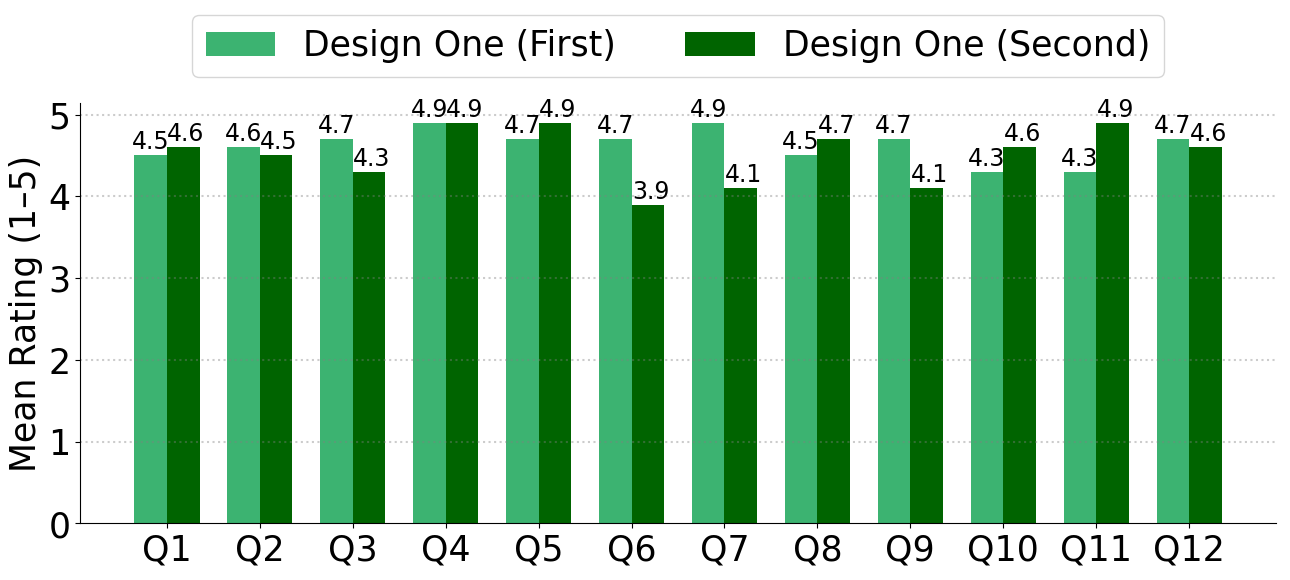}
\captionof{figure}{Design One survey results.}
\label{fig:designone}
\end{center}

In contrast, \textit{Design Two} performed worse overall, with noticeable effects based on the order of exposure. When presented first, it received an average score of 3.4 across all questions. However, when experienced after \textit{Design One}, the average rating of Group A dropped significantly to 2.9, indicating that prior exposure to a better-designed interface made users more aware of the flaws in the second design. As shown in Figure~\ref{fig:designtwo}, the most significant drops were observed in Q2 (Comfort), Q3 (Ease of Interaction), and Q4 (Organization of Elements), suggesting that users found \textit{Design Two} notably less comfortable, harder to use, and poorly structured when compared directly to a well-designed interface.

\begin{center}
\includegraphics[width=1\columnwidth]{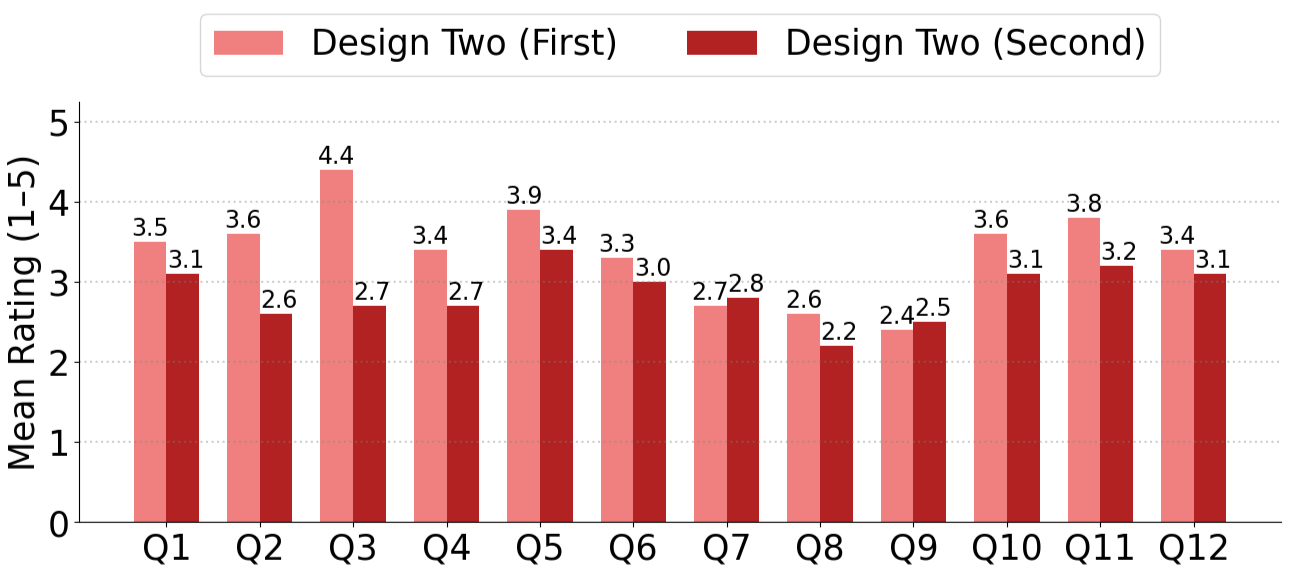}
\captionof{figure}{Design Two survey results.}
\label{fig:designtwo}
\end{center}

Mean survey scores (1–5) by question showing that Design One stays consistently high regardless of whether users saw it first or second, whereas Design Two’s scores fall significantly, particularly for comfort (Q2), interaction ease (Q3), and element organization (Q4), when experienced after the superior design.

\subsection{Overall Preference}
At the end of the study, participants were asked to choose their preferred design. 18 out of 20 participants (90\%) indicated a preference for \textit{Design One}, reinforcing the quantitative findings.

\subsection{Qualitative Feedback}
In addition to the questions, participants were given the option to provide open-ended feedback after using each design. This data offered valuable insight into the user experience that could not be fully captured by the survey.

Feedback for \textit{Design One} was predominantly positive. Participants frequently described the interface as intuitive, well-organized, easy to use, and enjoyable. One participant commented that the headset itself felt uncomfortable to wear, although this was not directly related to the interface design. A few remarks were made about the use of controllers. One right-handed participant stated, “The left controller was not necessary,” while another noted, “For scrolling, I prefer that the joystick input should not be inverted. This way it's more intuitive,” referring to the direction of the scroll action. 

The haptic feedback received mixed reactions. Some participants enjoyed the vibration, while others either did not notice it or did not like it. After comparing it with \textit{Design Two}, most users expressed a clear preference for \textit{Design One}. One participant observed, “Great consistent colors. The spacing was very good, there was good contrast, unlike the previous design.” However, a few mentioned that \textit{Design One} felt less colorful or visually unique compared to \textit{Design Two}.

In contrast, \textit{Design Two} received more critical feedback. Participants described the interface as unpleasant to use, unintuitive, and visually distracting. One participant summed it up as, “Functional but with very bad UI and UX, not pleasant to use.” However, another noted, “At the beginning it was a bit difficult to interact with, but after it becomes very intuitive and easy to use.” Color and readability issues were a recurring theme. One user said, “The color of the role field was too bright,” and another noted, “There were bad color choices like gray background for white text, and the colors weren’t synchronized with each other.” However, not all participants agreed. One mentioned that they preferred the font used in \textit{Design Two}, describing it as more interesting. 

Participants expressed mixed opinions about the feedback mechanisms in the interface. One user criticized the blurry save button in \textit{Design Two}, noting that it “did not work properly,” referring to the lack of visual feedback when it was pressed. Other participants, however, did not notice the lack of visual or haptic feedback, or felt that the feedback was sufficient due to the ray-cast visualization when interacting with the elements. 

Participants also commented on the positioning and distance of the interface. One participant suggested, “The possibility to drag the interface and center it from my point of view would be good.” Several users had to physically lean back to view the interface comfortably, as it was placed too close to their field of view. However, not all participants experienced this issue, possibly due to the short duration of the study, which may have prevented discomfort from becoming more noticeable over time.

There were mixed opinions on the preferred interaction method in both interface designs. Some participants favored using their hands, describing the experience as more fun and comfortable, while others preferred the controllers, saying they were more stable and sensitive to input. One participant suggested that a better approach might be to combine elements from both designs, like the menu design from \textit{Design One} with the hand-tracking from \textit{Design Two}. Another noted, “You can adjust the speed to match the controller one. It felt like I should adjust my hand speed in the beginning to get used to it.”

Overall, the qualitative feedback supports the survey findings. \textit{Design One} was perceived as intuitive and comfortable, while \textit{Design Two} suffered from inconsistent feedback and visual appeal, especially when evaluated after \textit{Design One}. However, participants’ interest in hand-tracking suggests that improving its implementation could significantly enhance future VR UI experiences.

\section{Discussion}\label{D}
The findings from this study highlight the effectiveness of the synthesized VR UI design guidelines and demonstrate their practical value in enhancing the user experience. The user study results clearly show that \textit{Design One}, which incorporated the best practices derived from the comprehensive literature review, significantly outperformed \textit{Design Two} across all survey questions. Participants rated \textit{Design One} 46\% higher (4.57 vs. 3.13 out of 5). These findings emphasize the importance of adhering to design principles that are specifically tailored for VR experiences.

The most noticeable improvements were observed in areas directly addressed by the guidelines: comfort, organization of interface elements, visual appeal, viewing distance, and color choices. These results align with the literature's emphasis on ergonomic design and spatial arrangement in VR environments. The strong user preference for \textit{Design One} (chosen by 90\% of participants) highlights the effectiveness of the 28 synthesized guidelines in meeting user needs and expectations. The four-level integration framework was particularly useful. Guidelines classified as 'Full Integration', including comfortable viewing distances, proper element spacing, and intuitive and ergonomic interactions, had the most significant impact on user experience.

The survey revealed that exposure order had an effect on user opinion. \textit{Design Two's }ratings dropped significantly when experienced after \textit{Design One} (from 3.4 to 2.9), while \textit{Design One} mostly maintained consistently high ratings regardless of the sequence. This suggests that exposure to well-designed interfaces increases users' awareness of poor design elements, making them more critical of suboptimal implementations. These results emphasize the necessity of implementing good design practices to form good user experiences.

Qualitative feedback supported these findings, with participants describing \textit{Design One} as intuitive and well-organized, while characterizing \textit{Design Two} as unpleasant and visually distracting.
While interaction methods (controller vs. hand-tracking) led to mixed user preferences, interface design quality had a greater impact on overall user experience. Visual design, spatial arrangement, and ergonomic considerations were identified as the most critical factors, with participants consistently criticizing \textit{Design Two's} poor color choices, inadequate contrast, and uncomfortable positioning.

The research successfully bridges the gap between theoretical research and practical application, providing concrete guidelines in an open-source repository that can be implemented by VR designers and developers. The \textit{FlUId} application serves as both a validation tool and an educational resource, demonstrating that these guidelines can be effectively translated into functioning examples. The clear user preference for interfaces that comply with the guidelines demonstrates the value of VR-specific design and confirms that good UI design can significantly improve the virtual experience.

\section{Future Work}\label{FW}
While this research has successfully established a comprehensive framework of VR UI design guidelines and validated their effectiveness through the \textit{FlUId} application, several opportunities exist for extending and refining this work.

An immediate next step is to integrate the settings menu design from this study into a full-scale VR application. This would allow the guidelines to be tested in a more complex, real-world environment where users interact with the interface as part of their primary task, rather than as an isolated evaluation. To assess the design’s effectiveness over time, extended exposure studies could be conducted to identify potential long-term comfort and usability issues. Such integration could reveal new challenges and considerations for designing interfaces in VR.

Another promising direction involves incorporating emerging technologies such as eye-tracking and advanced haptic feedback systems. These technologies offer new opportunities to enhance interaction, but also require the development of guidelines tailored to their unique capabilities. While such implementations were not feasible with the hardware available in this study, future research could focus on identifying design principles for gaze-based and haptic interaction across different VR platforms.

Finally, the development of practical development tools to support guideline implementation would significantly enhance their real-world use. These could include interface evaluation tools, template libraries, or integration plugins for widely used VR development platforms like Unity.

The foundation laid by this research provides a strong starting point for these future directions, with the potential to further improve user experience across a broad range of VR applications.

\section{Conclusion}\label{C}
This research addresses the lack of unified virtual reality user interface (VR-UI) design guidelines by synthesizing existing literature into a set of best practices and validating them through a user study. It introduces a structured framework for VR UI design integration that helps prioritize implementation based on usability impact. Through the development and evaluation of the \textit{FlUId} VR application, the study demonstrates that applying these guidelines leads to significantly better user experiences. 

Key design factors such as visual layout, spatial arrangement, and ergonomics were shown to be particularly important. The findings offer clear, practical direction for VR designers and developers, highlighting the value of tailored design approaches over traditional 2D adaptations. This work lays a strong foundation for improving VR usability and accessibility and contributes to the continuous advancement of immersive technology across various domains.

\section*{Aknowledgements}

This research was financially supported by the TUM Campus Heilbronn \textit{Incentive Fund 2024} of the Technical University of Munich, TUM Campus Heilbronn. We gratefully acknowledge their support, which provided the essential resources and opportunities to conduct this study. 

\printcredits

\bibliographystyle{cas-model2-names}
\bibliography{00Paper}

@article{gonzalez202310,
  title={10 Best practices in Immersive Learning Design and 10 points of connection with the Metaverse: a point of view},
  author={Gonzalez-Argote, Javier and Gonzalez-Argote, Denis},
  journal={Metaverse Basic and Applied Research},
  volume={2},
  pages={7--7},
  year={2023}
}

@article{Xie,
author = {Xie, Hanqing},
year = {2023},
month = {04},
pages = {189-198},
title = {The Applications of Interface Design and User Experience in Virtual Reality},
volume = {44},
journal = {Highlights in Science, Engineering and Technology},
}

@inproceedings{zhou2023design,
  title={Design paradigms of 3D user interfaces for VR exhibitions},
  author={Zhou, Yunzhan and Shi, Lei and He, Zexi and Li, Zhaoxing and Wang, Jindi},
  booktitle={IFIP Conference on Human-Computer Interaction},
  pages={618--627},
  year={2023},
  organization={Springer}
}

@article{wang2021authenticity,
  title={Authenticity, interactivity, and collaboration in virtual reality games: Best practices and lessons learned},
  author={Wang, Annie and Thompson, Meredith and Uz-Bilgin, Cigdem and Klopfer, Eric},
  journal={Frontiers in Virtual Reality},
  volume={2},
  pages={734083},
  year={2021},
  publisher={Frontiers Media SA}
}

@article{larocco2020developing,
  title={Developing the ‘best practices’ of virtual reality design: Industry standards at the frontier of emerging media},
  author={LaRocco, Michael},
  journal={Journal of visual culture},
  volume={19},
  number={1},
  pages={96--111},
  year={2020},
  publisher={SAGE Publications Sage UK: London, England}
}

@inproceedings{alves2020evaluation,
  title={Evaluation of graphical user interfaces guidelines for virtual reality games},
  author={Alves, Samuel and Callado, Arthur and Juc{\'a}, Paulyne},
  booktitle={2020 19th Brazilian Symposium on Computer Games and Digital Entertainment (SBGames)},
  pages={71--79},
  year={2020},
  organization={IEEE}
}

@article{ccatak2020guideline,
  title={A guideline study for designing virtual reality games},
  author={{\c{C}}atak, G{\"u}ven and Masalc{\i}, Server and {\c{S}}enyer, Seray},
  journal={AJIT-e: Academic Journal of Information Technology},
  volume={11},
  number={43},
  pages={12--36},
  year={2020},
  publisher={Akademik Bili{\c{s}}im Ara{\c{s}}t{\i}rmalar{\i} Derne{\u{g}}i}
}

@article{kelly2022design,
  title={Design considerations for supporting mindfulness in virtual reality},
  author={Kelly, Ryan M and Seabrook, Elizabeth M and Foley, Fiona and Thomas, Neil and Nedeljkovic, Maja and Wadley, Greg},
  journal={Frontiers in Virtual Reality},
  volume={2},
  pages={672556},
  year={2022},
  publisher={Frontiers Media SA}
}

@article{ramaseri2022systematic,
  title={A systematic survey on cybersickness in virtual environments},
  author={Ramaseri Chandra, Ananth N and El Jamiy, Fatima and Reza, Hassan},
  journal={Computers},
  volume={11},
  number={4},
  pages={51},
  year={2022},
  publisher={MDPI}
}

@article{velivckovic2021improvement,
  title={Improvement of the interaction model aimed to reduce the negative effects of cybersickness in VR rehab applications},
  author={Veli{\v{c}}kovi{\'c}, Predrag and Milovanovi{\'c}, Milo{\v{s}}},
  journal={Sensors},
  volume={21},
  number={2},
  pages={321},
  year={2021},
  publisher={MDPI}
}

@article{yahaya2022comparative,
  title={A Comparative Analysis on Cybersickness Reduction Guidelines in VR and IVR Applications for Children Road Safety Education.},
  author={Yahaya, Nur Sauri and Mutalib, Ariffin Abdul and Salam, Sobihatun Nur Abdul},
  journal={International Journal of Interactive Mobile Technologies},
  volume={16},
  number={5},
  year={2022}
}

@inproceedings{zhang2025inclusive,
  title={Inclusive avatar guidelines for people with disabilities: Supporting disability representation in social virtual reality},
  author={Zhang, Kexin and Spencer Jr, Edward Glenn Scott and Manikandan, Abijith and Li, Andric and Li, Ang and Yao, Yaxing and Zhao, Yuhang},
  booktitle={Proceedings of the 2025 CHI Conference on Human Factors in Computing Systems},
  pages={1--26},
  year={2025}
}

@article{ha2022novel,
  title={Novel hybrid brain-computer interface for virtual reality applications using steady-state visual-evoked potential-based brain--computer interface and electrooculogram-based eye tracking for increased information transfer rate},
  author={Ha, Jisoo and Park, Seonghun and Im, Chang-Hwan},
  journal={Frontiers in neuroinformatics},
  volume={16},
  pages={758537},
  year={2022},
  publisher={Frontiers Media SA}
}

@article{ijaz2022design,
  title={Design considerations for immersive virtual reality applications for older adults: a scoping review},
  author={Ijaz, Kiran and Tran, Tram Thi Minh and Kocaballi, Ahmet Baki and Calvo, Rafael A and Berkovsky, Shlomo and Ahmadpour, Naseem},
  journal={Multimodal technologies and interaction},
  volume={6},
  number={7},
  pages={60},
  year={2022},
  publisher={MDPI}
}

@article{kalaiselvi,
author = {Kalaiselvi, K and Bhuvaneswari, R. and Rekha, R. and Kumar, T.Rajesh and Banerjee, Rishav and Baheti, Omang},
year = {2024},
month = {03},
pages = {1056},
title = {VR based gesture elicitation for user—Interfaces with low vision},
volume = {7},
journal = {Journal of Autonomous Intelligence},
}

@article{bermejo2021exploring,
  title={Exploring button designs for mid-air interaction in virtual reality: A hexa-metric evaluation of key representations and multi-modal cues},
  author={Bermejo, Carlos and Lee, Lik Hang and Chojecki, Paul and Przewozny, David and Hui, Pan},
  journal={Proceedings of the ACM on Human-Computer Interaction},
  volume={5},
  number={EICS},
  pages={1--26},
  year={2021},
  publisher={ACM New York, NY, USA}
}

@article{caputo2019gestural,
  title={Gestural interaction in Virtual Environments: user studies and applications},
  author={Caputo, Ariel and others},
  year={2019}
}

@inproceedings{kaur2019exploring,
  title={Exploring 3D Interactions for Number Entry and Menu Selection in Virtual Reality Environment},
  author={Kaur, Akriti and Agrawal, Ashutosh and Yammiyavar, Pradeep},
  booktitle={Research into Design for a Connected World: Proceedings of ICoRD 2019 Volume 2},
  pages={781--791},
  year={2019},
  organization={Springer}
}

@inproceedings{regazzoni2018virtual,
  title={Virtual reality applications: guidelines to design natural user interface},
  author={Regazzoni, Daniele and Rizzi, Caterina and Vitali, Andrea},
  booktitle={International Design Engineering Technical Conferences and Computers and Information in Engineering Conference},
  volume={51739},
  pages={V01BT02A029},
  year={2018},
  organization={American Society of Mechanical Engineers}
}

@inproceedings{cardona2020natural,
  title={Natural Interfaces to Support ADHD in Virtual Reality Environments},
  author={Cardona-Reyes, H{\'e}ctor and Barba-Gonzalez, Maria Lorena and Mu{\~n}oz-Arteaga, Jaime and Gonzalez-Romo, Ivan and Alvarez-Rodriguez, Francisco},
  booktitle={2020 3rd International Conference of Inclusive Technology and Education (CONTIE)},
  pages={20--27},
  year={2020},
  organization={IEEE}
}

@article{kaur2020cognitive,
  title={A cognitive load assessment study of three-dimensional interactive virtual reality interfaces},
  author={Kaur, Akriti and Agrawal, Mudit and Yammiyavar, Pradeep G},
  journal={International Journal of Forensic Engineering and Management},
  volume={1},
  number={1},
  pages={103--115},
  year={2020},
  publisher={Inderscience Publishers (IEL)}
}

@inproceedings{kaur2017comparative,
  title={A comparative study of 2D and 3D mobile keypad user interaction preferences in virtual reality graphic user interfaces},
  author={Kaur, Akriti and Yammiyavar, Pradeep G},
  booktitle={Proceedings of the 23rd ACM symposium on virtual reality software and technology},
  pages={1--2},
  year={2017}
}

@inproceedings{azai2018open,
  title={Open palm menu: A virtual menu placed in front of the palm},
  author={Azai, Takumi and Otsuki, Mai and Shibata, Fumihisa and Kimura, Asako},
  booktitle={Proceedings of the 9th Augmented Human International Conference},
  pages={1--5},
  year={2018}
}

@inproceedings{azai2018tap,
  title={Tap-tap menu: body touching for virtual interactive menus},
  author={Azai, Takumi and Ushiro, Syunsuke and Li, Junlin and Otsuki, Mai and Shibata, Fumihisa and Kimura, Asako},
  booktitle={Proceedings of the 24th ACM Symposium on Virtual Reality Software and Technology},
  pages={1--2},
  year={2018}
}

@incollection{kwon2017spatial,
  title={A spatial user interface design using accordion metaphor for VR systems},
  author={Kwon, JoungHuem and Kim, YoungEun and Nam, SangHun},
  booktitle={SIGGRAPH Asia 2017 Posters},
  pages={1--2},
  year={2017}
}

@inproceedings{pandey2024pilot,
  title={A Pilot Study on Hand-Referenced Menu User Interfaces for Head-Mounted Display Virtual Reality},
  author={Pandey, Rishabh and Sorathia, Keyur},
  booktitle={Proceedings of the 27th International Academic Mindtrek Conference},
  pages={260--263},
  year={2024}
}

@article{kim2022pseudo,
  title={Pseudo-haptic button for improving user experience of mid-air interaction in VR},
  author={Kim, Woojoo and Xiong, Shuping},
  journal={International Journal of Human-Computer Studies},
  volume={168},
  pages={102907},
  year={2022},
  publisher={Elsevier}
}

@inproceedings{dudley2019performance,
  title={Performance envelopes of virtual keyboard text input strategies in virtual reality},
  author={Dudley, John and Benko, Hrvoje and Wigdor, Daniel and Kristensson, Per Ola},
  booktitle={2019 IEEE International Symposium on Mixed and Augmented Reality (ISMAR)},
  pages={289--300},
  year={2019},
  organization={IEEE}
}

@inproceedings{dingler2018vr,
  title={Vr reading uis: Assessing text parameters for reading in vr},
  author={Dingler, Tilman and Kunze, Kai and Outram, Benjamin},
  booktitle={Extended Abstracts of the 2018 CHI Conference on Human Factors in Computing Systems},
  pages={1--6},
  year={2018}
}

@article{laine2024investigating,
  title={Investigating User Experience of an Immersive Virtual Reality Simulation Based on a Gesture-Based User Interface},
  author={Laine, Teemu H and Suk, Hae Jung},
  journal={Applied Sciences},
  volume={14},
  number={11},
  pages={4935},
  year={2024},
  publisher={MDPI}
}

@article{byam2019best,
  title={Best practices and metrics for virtual reality user interfaces},
  author={Byam, Jay},
  year={2019}
}

@article{kim2017study,
  title={A study on interaction of gaze pointer-based user interface in mobile virtual reality environment},
  author={Kim, Mingyu and Lee, Jiwon and Jeon, Changyu and Kim, Jinmo},
  journal={Symmetry},
  volume={9},
  number={9},
  pages={189},
  year={2017},
  publisher={MDPI}
}

@article{ap2018movement,
  title={Movement modalities in virtual reality: a case study from ocean rift examining the best practices in accessibility, comfort, and immersion},
  author={ap Cenydd, Llyr and Headleand, Christopher J},
  journal={IEEE Consumer Electronics Magazine},
  volume={8},
  number={1},
  pages={30--35},
  year={2018},
  publisher={IEEE}
}

@inproceedings{otaran2024exploring,
  title={Exploring User Preferences for Walking in Virtual Reality Interfaces Through an Online Questionnaire},
  author={Otaran, Ata and Farkhatdinov, Ildar},
  booktitle={International Conference on Human-Computer Interaction},
  pages={244--258},
  year={2024},
  organization={Springer}
}

@article{bektacs2021systematic,
  title={The systematic evaluation of an embodied control interface for virtual reality},
  author={Bekta{\c{s}}, Kenan and Thrash, Tyler and van Raai, Mark A and K{\"u}nzler, Patrik and Hahnloser, Richard},
  journal={Plos one},
  volume={16},
  number={12},
  pages={e0259977},
  year={2021},
  publisher={Public Library of Science San Francisco, CA USA}
}

@article{hamdanindustry,
  title={From Industry 4.0 to Industry 5.0},
  author={Hamdan, Allam and Harraf, Arezou and Buallay, Amina and Arora, Pallvi and Alsabatin, Hala},
  publisher={Springer}
}

@inproceedings{saranya2024development,
  title={Development of Virtual Reality Platform through Human Computer Interaction using Artificial Intelligence},
  author={Saranya, S and Channarayapriya, B and Harshavardhini, U and Nandhini, A Sunitha and Revathi, J and Venkatesan, R},
  booktitle={2024 3rd International Conference on Applied Artificial Intelligence and Computing (ICAAIC)},
  pages={283--288},
  year={2024},
  organization={IEEE}
}

@inproceedings{puspitasari2022review,
  title={Review of Persuasive User Interface as strategy for technology addiction in Virtual Environments},
  author={Puspitasari, Fachrina Dewi and Lee, Lik-Hang},
  booktitle={2022 IEEE International Symposium on Mixed and Augmented Reality Adjunct (ISMAR-Adjunct)},
  pages={44--54},
  year={2022},
  organization={IEEE}
}

@article{chen2024survey,
  title={A Survey on the Design of Virtual Reality Interaction Interfaces},
  author={Chen, Meng-Xi and Hu, Huicong and Yao, Ruiqi and Qiu, Longhu and Li, Dongxu},
  journal={Sensors},
  volume={24},
  number={19},
  pages={6204},
  year={2024},
  publisher={MDPI}
}

@article{yeo2024entering,
  title={Entering the next dimension: A review of 3d user interfaces for virtual reality},
  author={Yeo, Adriel and Kwok, Benjamin WJ and Joshna, Angelene and Chen, Kan and Lee, Jeannie SA},
  journal={Electronics},
  volume={13},
  number={3},
  pages={600},
  year={2024},
  publisher={MDPI}
}

@article{kharoub20193d,
  title={3D user interface design and usability for immersive VR},
  author={Kharoub, Hind and Lataifeh, Mohammed and Ahmed, Naveed},
  journal={Applied sciences},
  volume={9},
  number={22},
  pages={4861},
  year={2019},
  publisher={MDPI}
}

@article{khanvirtual,
  title={Virtual Reality (VR) User Interfaces: Guidelines for Human Factors and Ergonomic Design},
  author={Khan, Zainab and Chellappa, Vigneshkumar and Ginda, Grzegorz},
 publisher={IEOM Society International},
year={2024}
}

@inproceedings{kojic2020user,
  title={User experience of reading in virtual reality—finding values for text distance, size and contrast},
  author={Koji{\'c}, Tanja and Ali, Danish and Greinacher, Robert and M{\"o}ller, Sebastian and Voigt-Antons, Jan-Niklas},
  booktitle={2020 Twelfth International Conference on Quality of Multimedia Experience (QoMEX)},
  pages={1--6},
  year={2020},
  organization={IEEE}
}

@article{wang2021experimental,
  title={An experimental investigation of menu selection for immersive virtual environments: Fixed versus handheld menus},
  author={Wang, Yanbin and Hu, Yizhou and Chen, Yu},
  journal={Virtual Reality},
  volume={25},
  number={2},
  pages={409--419},
  year={2021},
  publisher={Springer}
}

@inproceedings{chandana2023exploring,
  title={Exploring the frontiers of user experience design: Vr, ar, and the future of interaction},
  author={Chandana, B Hari and Shaik, Nazeer and Chitralingappa, P},
  booktitle={2023 International Conference on Computer Science and Emerging Technologies (CSET)},
  pages={1--6},
  year={2023},
  organization={IEEE}
}

@inproceedings{lima2024construction,
  title={Construction of Interfaces in Virtual Reality: a systematic review focusing on Ergonomics and User Experience},
  author={Lima, Sofia and Catapan, M{\'a}rcio Fontana and Sasaki Zeredo, Jun},
  booktitle={Proceedings of the 26th Symposium on Virtual and Augmented Reality},
  pages={257--260},
  year={2024}
}

@article{shoikova2017best,
  title={Best practices for designing user experience for Internet of Things and virtual reality},
  author={Shoikova, Elena and Peshev, Anatoly},
  journal={Electrotechnica \& Electronica (E+ E)},
  volume={52},
  year={2017}
}

@book{wall2021empirical,
  title={An empirical study of virtual reality menu interaction and design},
  author={Wall, Emily Salmon},
  year={2021},
  publisher={Mississippi State University}
}

@inproceedings{dombrowski2019designing,
  title={Designing inclusive virtual reality experiences},
  author={Dombrowski, Matt and Smith, Peter A and Manero, Albert and Sparkman, John},
  booktitle={International Conference on Human-Computer Interaction},
  pages={33--43},
  year={2019},
  organization={Springer}
}

@article{Anuyahong_Pengnate_2023, 
title={Exploring the Impact of User Interface Design on User Experience in Unity VR Games}, 
volume={11}, 
number={4}, 
journal={International Journal of Scientific Research in Computer Science and Engineering}, 
author={Anuyahong, Bundit and Pengnate, Wipanee}, 
year={2023}, 
month={Aug.}, 
pages={24–30} 
}

@inproceedings{chauvergne2023user,
  title={User onboarding in virtual reality: An investigation of current practices},
  author={Chauvergne, Edwige and Hachet, Martin and Prouzeau, Arnaud},
  booktitle={Proceedings of the 2023 CHI conference on human factors in computing systems},
  pages={1--15},
  year={2023}
}

@article{li2024investigating,
  title={Investigating Creation Perspectives and Icon Placement Preferences for On-Body Menus in Virtual Reality},
  author={Li, Xiang and He, Wei and Jin, Shan and Gugenheimer, Jan and Hui, Pan and Liang, Hai-Ning and Kristensson, Per Ola},
  journal={Proceedings of the ACM on Human-Computer Interaction},
  volume={8},
  number={ISS},
  pages={236--254},
  year={2024},
  publisher={ACM New York, NY, USA}
}

@article{lou2021hand,
  title={Hand-adaptive user interface: improved gestural interaction in virtual reality},
  author={Lou, Xiaolong and Li, Xiangdong A and Hansen, Preben and Du, Peng},
  journal={Virtual Reality},
  volume={25},
  pages={367--382},
  year={2021},
  publisher={Springer}
}

@inproceedings{lediaeva2020evaluation,
  title={Evaluation of body-referenced graphical menus in virtual environments},
  author={Lediaeva, Irina and LaViola, Joseph},
  booktitle={Graphics Interface 2020},
  year={2020}
}

@article{wentzel2024comparison,
  title={A Comparison of Virtual Reality Menu Archetypes: Raycasting, Direct Input, and Marking Menus},
  author={Wentzel, Johann and Lakier, Matthew and Hartmann, Jeremy and Shazib, Falah and Casiez, G{\'e}ry and Vogel, Daniel},
  journal={IEEE Transactions on Visualization and Computer Graphics},
  year={2024},
  publisher={IEEE}
}

@inproceedings{berrezuetaguzman2025,
  title={A therapeutic role-playing VR game for children with intellectual disabilities},
  author={Berrezueta-Guzman, Santiago and Chen, WenChun and Wagner, Stefan},
  booktitle={2025 International Conference on Metaverse Computing, Networking and Applications (MetaCom)},
  pages={389--396},
  year={2025},
  organization={IEEE}
}

@article{sobchyshak2025,
  title={Pushing the boundaries of immersion and storytelling: A technical review of unreal engine},
  author={Sobchyshak, Oleksandra and Berrezueta-Guzman, Santiago and Wagner, Stefan},
  journal={Displays},
  pages={103268},
  year={2025},
  publisher={Elsevier}
}

\end{multicols}
\end{document}